\documentclass[superscriptaddress]{article}
\usepackage{amssymb,amsthm,amsmath,amsfonts,physics,algorithm2e,float}
\usepackage[a4paper, total={7in, 9in}]{geometry}
\usepackage{ulem,soul,enumerate}
\usepackage{graphicx}
\usepackage[pdftex,dvipsnames,usenames]{xcolor}
\usepackage[colorlinks=true,urlcolor=blue,citecolor=blue,linkcolor=blue]{hyperref}
\usepackage{authblk}

\usepackage{multicol}
\usepackage{caption}
\usepackage{subcaption}
\begin{document}

\title{Experimental Multi-state Quantum Discrimination in the Frequency Domain with Quantum Dot Light}

\author[1]{Alessandro Laneve\thanks{alessandro.laneve@uniroma1.it}}
\author[1]{Michele B. Rota}
\author[1]{Francesco Basso Basset}
\author[1]{Nicola P. Fiorente}
\author[2]{Tobias M. Krieger}
\author[2]{Saimon F. Covre da Silva}
\author[3]{Quirin Buchinger}
\author[4]{Sandra Stroj}
\author[3]{Sven Hoefling}
\author[3]{Tobias Huber-Loyola}
\author[2]{Armando Rastelli}
\author[1]{Rinaldo Trotta}
\author[1]{Paolo Mataloni}

\affil[1]{Dipartimento di Fisica, Sapienza Universit\`{a} di Roma, Piazzale Aldo Moro, 5, I-00185 Roma, Italy}
\affil[2]{Institute of Semiconductor and Solid State Physics, Johannes Kepler University, Altenbergerstraße 69, Linz 4040, Austria}
\affil[3]{Technical Physics, University of Wuerzburg, Am Hubland, D-97074 Wuerzburg, Germany}
\affil[4]{Forschungszentrum
Mikrotechnik, FH Vorarlberg, Hochschulstr. 1, A-6850 Dornbirn, Austria.}

\date{\today}

\maketitle

\begin{abstract}
    The quest for the realization of effective quantum state discrimination strategies is of great interest for quantum information technology, as well as for fundamental studies. Therefore, it is crucial to develop new and more efficient methods to implement discrimination protocols for quantum states. Among the others, single photon implementations are more advisable, because of their inherent security advantage  in quantum communication scenarios.
    In this work, we present the experimental realization of a protocol employing a time-multiplexing strategy to optimally discriminate among eight non-orthogonal states, encoded in the four-dimensional Hilbert space spanning both the polarization degree of freedom and photon energy. The experiment, built on a custom-designed bulk optics analyser setup and single photons generated by a nearly deterministic solid-state source, represents a benchmarking example of minimum error discrimination with actual quantum states, requiring only linear optics and two photodetectors to be realized. Our work paves the way for more complex applications and delivers a novel approach towards high-dimensional quantum encoding and decoding operations. 
\end{abstract}

\begin{multicols}{2}
\section*{Introduction}
One of the fundamental and most exciting features of quantum systems consists of their ability to linger in a superposition of two orthogonal states. Because of this, a system described in terms of a 2D Hilbert space, spanned by two orthogonal vectors, can in principle assume an infinite number of different states. These states are not orthogonal, in general, and the number of projective measurements necessary to discriminate among them grows as the number of allowed states grows. If we do not bind in any way the states that the system can occupy, we would need to perform an infinite number of measurements to have a perfect identification of its state, hence we would need an infinite number of copies of the system.\\
This issue is particularly relevant in Quantum Communication scenarios: a single qubit features a potentially infinite space for encoding, but it is hard to retrieve this information. We would generally like to communicate using the largest possible alphabet, providing the highest possible rate of success, and saving resources, in particular the amount of signal and the number of projective measurements that have to be performed. (this latter point is also important in terms of communication security \cite{brassard2000limitations,beveratos2002single}).\\ This represents the main aim of Quantum State Discrimination (QSD), which featured many developments and applications in the last two decades, ranging from straightforward quantum communication problems \cite{chefles2000quantum,bae2015quantum}, to dimension witnessing \cite{brunner2013dimension}, to the discrimination of unitary operations \cite{acin2001statistical,pirandola2019fundamental}
and quantum foundations \cite{pusey2012reality, schmid2018contextual}.
Two main approaches to the problem of QSD can be considered: the first one, known as Minimum Error Discrimination (MED) \cite{bae2015quantum} or Quantum Hypothesis Testing \cite{helstrom1976quantum,chefles2000quantum}, is conceived to discriminate among a certain set of states featuring the minimum probability of error in guessing the state; the second one, known as Unambiguous State Discrimination (USD) \cite{chefles2000quantum,bae2015quantum,ivanovic1987differentiate}, aims at discriminating without errors among a given set of states, but admitting for inconclusive measurements.
Other hybrid approaches are possible, too \cite{holevo1973bounds,croke2006maximum,slussarenko2017quantum}.\\
The search for increasingly  more and more powerful QSD protocols is naturally linked to the pursuit of high-dimensional quantum information carriers. It is in general important for quantum communication applications that a single carrier can be filled with as much information as possible, i.e., that high-dimensional systems are harnessed \cite{cozzolino2019high}. Indeed, \textit{qudits} can carry much more information than both \textit{qubits} and classical bits \cite{cover1999elements}. As an instance, entangled \textit{qudits} have been demonstrated to beat the classical channel capacity \cite{barreiro2008beating,hu2018beating,dixon2012quantum}. Moreover, high-dimensional quantum states are quite robust to noise, either environmental or due to eavesdropping, granting a greater security in Quantum Communication \cite{sheridan2010security,cerf2002security,bechmann2000quantum}.
Thus, any increase in the dimensionality of systems employed for Quantum Information tasks can represent a step towards more powerful and secure Quantum Communication \cite{cozzolino2019high}.
In parallel, it is also important to develop Quantum State Discrimination protocols working with high-dimensional carriers, in order to increase their effectiveness and range of application.\\
The ideal platform for quantum communication, and QSD, is photonics. Single photons represent one of the best candidates for the role of flying qubits and feature several fundamental properties such as resistance to decoherence, ease of manipulation with bulk linear optic elements, possibility of encoding information in their several degrees of freedom \cite{flamini2018photonic}. 
In the wide landscape of single-photon emitters, semiconductor quantum dots (QDs) are emerging as promising candidate for the role. They can emit single and entangled photons on-demand \cite{michler2000quantum,jayakumar2013deterministic}, with high level of indistinguishability \cite{zhai2022quantum} and high degree of entanglement, ultra-low multiphoton emission probability \cite{schweickert2018demand} and their properties can be enhanced with the combination of photonic cavities \cite{liu2019solid,wang2019towards,wang2019demand,somaschi2016near} and strain-tuning \cite{huber2018strain}. Using quantum light from semiconductors QD already led to the demonstration of quantum communication protocols such as quantum teleportation \cite{nilsson2013quantum,reindl2018all} and entanglement swapping \cite{basset2019entanglement,zopf2019entanglement} and their use in quantum key distribution \cite{kupko2020tools, vajner2022quantum,basso2021quantum} 
Progress in the performance of QDs in terms of brightness and indistinguishability \cite{wang2019towards,tomm2021bright} are making these emitters ready for proof of concept applications in quantum cryptography. By exploiting multi-excitonic cascades in a single QD, one can achieve the emission of multiple photons with different energies. In fact, the emission of two photons with a different energy in the biexciton-exciton cascade has already been exploited for spectral multiplexing in quantum cryptography \cite{aichele2004separating}. This offers an example of a use case for increasing the Hilbert space dimension of a quantum state, combining the frequency degrees of freedom with single photon polarization.\\
In fact, light polarization has been the first and main degree of freedom exploited for experimental Quantum State Discrimination and the only one to be employed in actual single photon experimental attempts \cite{clarke2001unambiguous,clarke2001overcomplete,mohseni2004optical,solis2016experimental}.
In a coherent state experimental framework, it is possible to increase the system dimensionality in different ways, such as Quantum Phase Shift Keying (QPSK) \cite{becerra2011m,becerra2013experimental,becerra2013implementation,burenkov2020time, sidhu2021quantum}
or the employment of complex modal structures \cite{solis2017experimental}.
Otherwise, it is also possible to exploit multiple copies of the system to increase effectiveness of QSD protocols and allowing for adaptivity of measurement operations \cite{acin2005multiple,assalini2011revisiting,slussarenko2017quantum, becerra2013experimental,ferdinand2017multi}.\\
In the spirit of adaptive protocols, some efforts have been devoted to the development of dynamical decoding strategies relying on Quantum Walks \cite{kurzynski2013quantum,li2019implementation}, neural networks \cite{caruso2020qsd,patterson2021quantum} or general sequential measurement schemes \cite{bergou2013extracting,namkung2018analysis}.
In a previous experiment, some of the authors followed this dynamical approach, realizing an optical network for the MED of four non-orthogonal polarization states of single photons, relying on the pairing between detection time and input state \cite{laneve2022experimental}.\\
In this work, we exploit such time multiplexing method as a tool to achieve a benchmarking result in the framework of single-photon QSD protocols. We are able to implement an experimental MED protocol for the discrimination of eight non-orthogonal states, using single photons generated by a QD. We harness the photon energy as an additional degree of freedom to generate a 4D encoding space. Specifically, the eight states are encoded in the joint 4D Hilbert space obtained composing the photon polarization degree of freedom and the 2D space spanned by two different frequencies of the single photons. 
Our strategy allows discrimination among the eight states up to the minimum theoretical error probability, relying on an evolution scheme mapping both polarization and frequency in time-wise probability distributions.\\
By our experiment, photon energy is used as an additional encoding variable in Quantum State Discrimination problems for the first time. In general, frequency-based quantum information processing has been extensively investigated only in the last years
\cite{cozzolino2019high, kobayashi2017mach,lu2018electro,lu2019quantum}, thus leaving room for further research. 

\section*{Results}

\subsection*{Protocol description}

Our aim is the discrimination of eight non-orthogonal states encoded in a \textit{ququart} with the maximum probability of success.
Our approach focuses on saving resources, in terms of hardware and signal, and also on the retention of a fundamental-level security that can only be guaranteed by the employment of single photons.\\
Therefore, we want each measurement operation we perform to deliver a significant result, thus we consider the MED framework. We wish to perform discrimination among N states of a $D$-dimensional system, achieving the highest possible average probability of correct guess. We consider a set of geometrically uniform states (GUSs), i.e., a set of equally likely states featuring a particular symmetry with respect to a given transformation  $\hat{U}$; specifically, they form a finite Abelian commutative group \cite{eldar2001quantum}. We are taking into account the case of geometrically uniform \textit{quantum} states rather than their coherent QPSK counterpart. In any case, we know that for GUSs the minimum error probability of guess (averaged over the eight states) that is achievable is $P_{err}=1-\frac{D}{N}$ \cite{yuen1975optimum}.
In particular, we are interested in the $N=8$, $D=4$ case, which yields a minimum $P_{err}=0.5$, that is in turn equal to the successful guess maximum probability $P_{guess}=1-P_{err}=0.5$.\\
The space we consider is the composition of the one spanned by the orthogonal polarization basis ${\ket{H},\ket{V}}$ and the one spanned by a basis of two different frequencies $\ket{\omega_1},\ket{\omega_2}$. In principle, light frequency spans an infinite dimensional space, but we limit to two possible distinguishable frequencies, that we are able to prepare our input states in and that we are able to discriminate. Hence, we have a joint basis $\{\ket{H}\ket{\omega_1},\ket{V}\ket{\omega_1},\ket{H}\ket{\omega_2},\ket{V}\ket{\omega_2}\}$.\\
We choose eight states that are globally non-orthogonal, although some subsets of them are:

\begin{equation}
    \begin{cases}
    \ket{\psi_1}=\ket{+}\ket{\omega_1}\\
    \ket{\psi_2}=\ket{-}\ket{\omega_1}\\
    \ket{\psi_3}=\ket{H}\ket{\omega_1}\\
    \ket{\psi_4}=\ket{V}\ket{\omega_1}\\
    \ket{\psi_5}=\ket{+}\ket{\omega_2}\\
    \ket{\psi_6}=\ket{-}\ket{\omega_2}\\
    \ket{\psi_7}=\ket{H}\ket{\omega_2}\\
    \ket{\psi_8}=\ket{V}\ket{\omega_2}
    \end{cases}
    \label{eq:states}
\end{equation}

where $\ket{+}=\frac{\ket{H}+\ket{V}}{\sqrt{2}}$ and $\ket{-}=\frac{\ket{H}-\ket{V}}{\sqrt{2}}$.
This set is not actually a geometrical uniform set but, in practice, we can treat it as such, since it is composed of two orthogonal subsets that are, in turn, geometrically uniform.\\
Our discrimination strategy can be summarized as a four state discrimination analogous to the one in Ref.  \cite{laneve2022experimental}, followed by a binary discrimination among the two orthogonal frequency subspaces.

\begin{figure}[H]
\raggedleft
    \includegraphics[width=1.1\columnwidth]{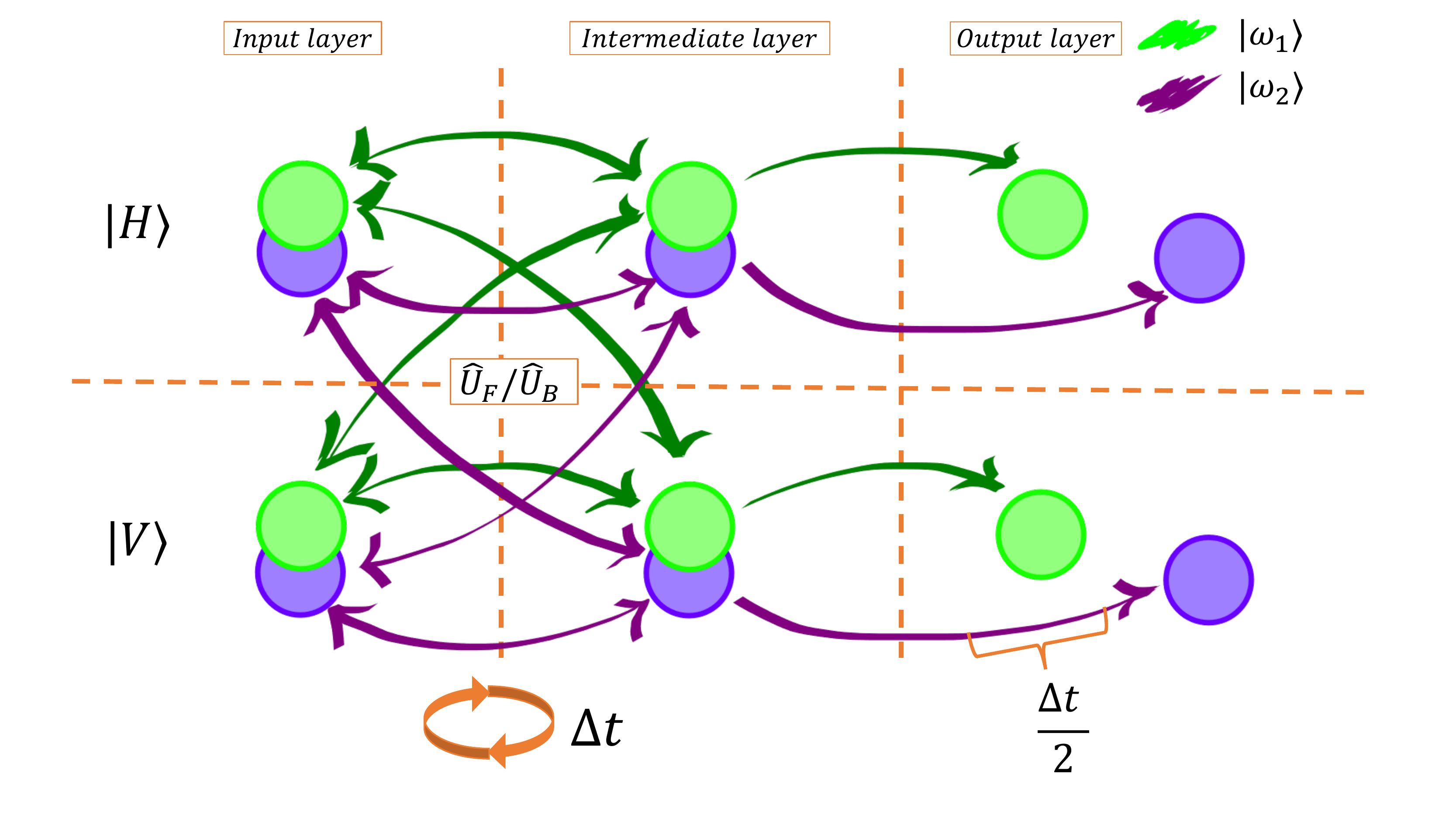}
    \caption{\textbf{Model of the network-like receiver.}\textit{ The system is prepared in a certain frequency and polarization state, that corresponds to a superposition state of input layer nodes. The system evolves forth and back from the input layer to the intermediate one of two parallel networks, according to the unitary operators $\hat{U}_F$ and $\hat{U}_B$, with a chance of being sent to the output when in the intermediate layer. There, the two identical networks become different, since the output with $\omega_2$ frequency is delayed with respect to the one featuring $\omega_1$. Specifically, this delay is half the travel time of the back and forth evolution between the intermediate and input layers. In this way, signal featuring one of the two frequencies is maximally distinguishable from the other in time.}}
    \label{fig:network}
\end{figure}

This is performed thanks to the imposition of a periodical dynamics on the polarization state of the system. This dynamics makes possible to associate the detection of a photon after a certain evolution step and in a certain polarization state with the presence of a specific input state among the possible eight given in Eq. \ref{eq:states}.\\
Our protocol can be described as the processing of the input state through the network depicted in Fig. \ref{fig:network}, inspired by \cite{dallapozza2020quantum}: the polarization state is sent forward and backward through the input and intermediate layers of the network, with a certain probability of being sent to the output layer. The frequency degree of freedom, on the other hand, doesn't change throughout the evolution and it is globally discriminated by a final projection. There is no way that the two frequencies are mistaken; therefore, states in different frequency subspaces can be regarded as processed through different networks.
The evolution from the input layer to the intermediate layer is mediated by the \textit{forward} unitary evolution $\hat{U}_F$, while the \textit{backward} unitary evolution from the intermediate layer back to the input one is $\hat{U}_B$. These are $2\times2$ unitary matrices acting on the polarization degree of freedom only. These operators act on both frequency subspaces in parallel. 
%
%
After the application of $\hat{U}_F$ the system is in the intermediate layer. With a certain probability, it is either sent to the output layer, where the frequency discrimination takes place, or it keeps evolving. \\
We now consider the polarization-wise discrimination: in the output layer, we can detect the system in one of two states of polarization, $\ket{H}$ or $\ket{V}$, that from now on we denote as $\ket{H}=\pi_1$ and $\ket{V}=\pi_2$. The probability of finding the system in $\pi_1$ or $\pi_2$ depends on the initial state and the dynamics it is subject to. The dynamics we impose is such that if the system does not propagate towards the output, it may return at a different evolution step after travelling back and forth through the network. At this point, the probability for the system to be found in one of the two polarization states may have changed. Therefore, we consider the conditional probabilities $P(\{t,\pi\}|\psi_i)$ and manipulate them by tailoring the evolution parameters. In particular, we can associate a specific duple $\{t_i,\pi_i\}$ to each state $\ket{\psi_i}$ and set the evolution unitary so as to maximize the probabilities $P(\{t_i,\pi_i\}|\psi_i)$. If the initial state is $\ket{\psi_i}$, the probability that the system is detected at the chosen time bin $t_i$ with chosen polarization $\pi_i$ is maximum.
This kind of strategy produces an optimal protocol for four geometrically uniform states, as demonstrated in \cite{laneve2022experimental}.\\
In order to apply this method to the eight states defined in Eq. \ref{eq:states}, we have to split the time bins we are considering: for each time bin $t_i$ we consider two sub-bins, one for each frequency. We apply to one of the two frequencies ($\ket{\omega_2}$ arbitrarily) a delay line, that provides a time delay for the detection of states in the $\ket{\omega_2}$ subspace which we set as half of the cycle time of the network. This is schematically depicted in Fig. \ref{fig:network}.
We can consider all the time bins and renumber them, so that detection in even time bins $\{t_0,t_2,t_4,...\}$ corresponds to an initial state in the $\ket{\omega_1}$ subspace while detection in odd time bins $\{t_1,t_3,t_5,...\}$ reveals that the system features an $\ket{\omega_2}$ frequency.
Thanks to this strategy, we can in principle produce a discrimination protocol for eight polarization-frequency states with the optimal probability of correct guess $P_{guess}=0.5$.

\subsection*{Experimental realization}

\begin{figure*}

       \centering
        \includegraphics[width=1.6\columnwidth]{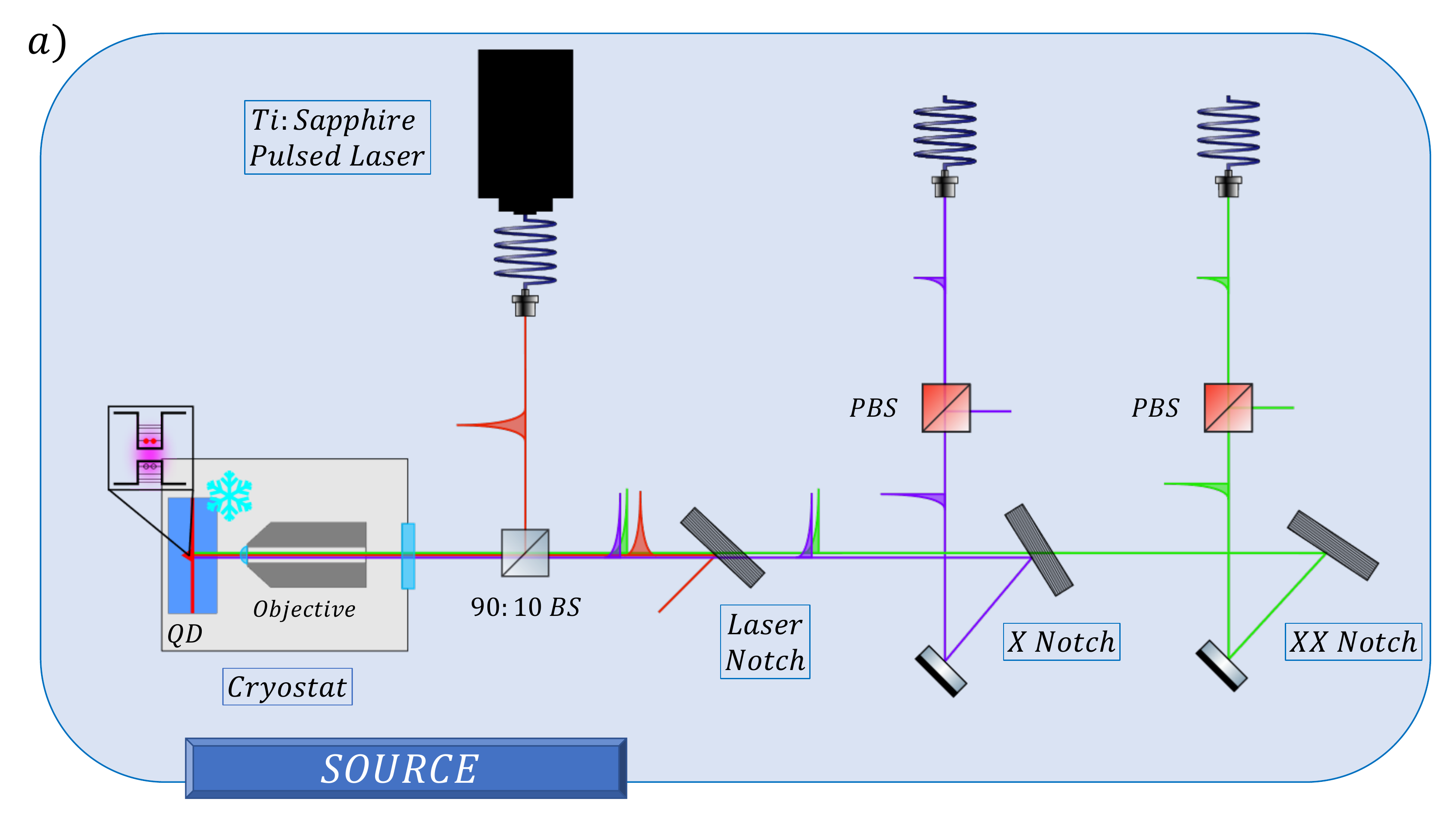}
         \includegraphics[width=1.6\columnwidth]{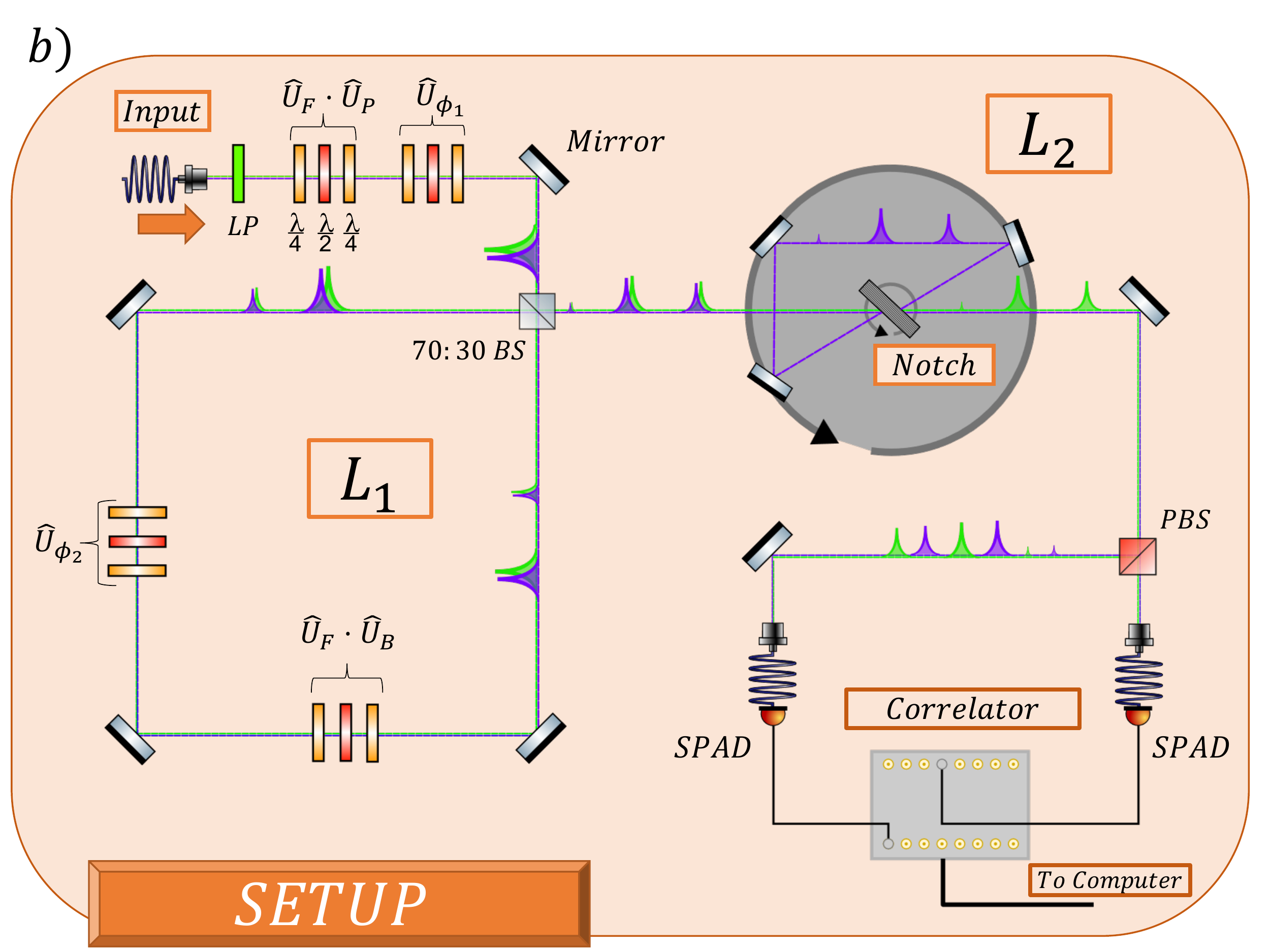}
        \caption{\textbf{Experimental scheme of the a) source setup and the b) optical receiver.} \textit{A Ti:Sa Pulsed Laser pumps a Quantum Dot (QD), that generates single photons at different frequencies. 
        The residual laser light is filtered out by an optical volume Bragg grating (VBG) filter suitably oriented.
        The $\ket{H}$ output of the exciton (purple) or the biexciton (green) light is used as an input for the b) actual discriminator, depending on the state we wish to prepare. Polarization state is prepared through a Linear Polarizer (LP) and a waveplates set QWP-HWP-QWP ($\frac{\lambda}{4}-\frac{\lambda}{2}-\frac{\lambda}{4}$ in the picture) that also drive the first evolution step, globally applying $\hat{U}_F\cdot\hat{U}_P$.  Along the loop $L_1$, a set of waveplates imposes the transformation $\hat{U}_F\cdot\hat{U}_B$ on photon polarization. Further sets are employed for phase compensation through the unitaries $\hat{U}_{\phi_1}$, $\hat{U}_{\phi_2}$. After escaping $L_1$, the signal arrives on a VBG set to reflect photons at the exciton frequency $\omega_2$, sending them in a second optical loop $L_2$. The signal is reflected again, after travelling for a length $l_{L_2}=l_{L_1}/2$, so that $\omega_1$ and $\omega_2$ are time delayed and perfectly distinguishable. Eventually, photons are coupled to multimode fibers and sent to Single Photon Avalanche Diodes (SPADs), connected to a correlator and interfaced to a computer.
        } }
        \label{fig:setup}
\end{figure*}

\begin{figure*}
       \begin{subfigure}{0.5\textwidth}
       \includegraphics[width=1.\columnwidth]{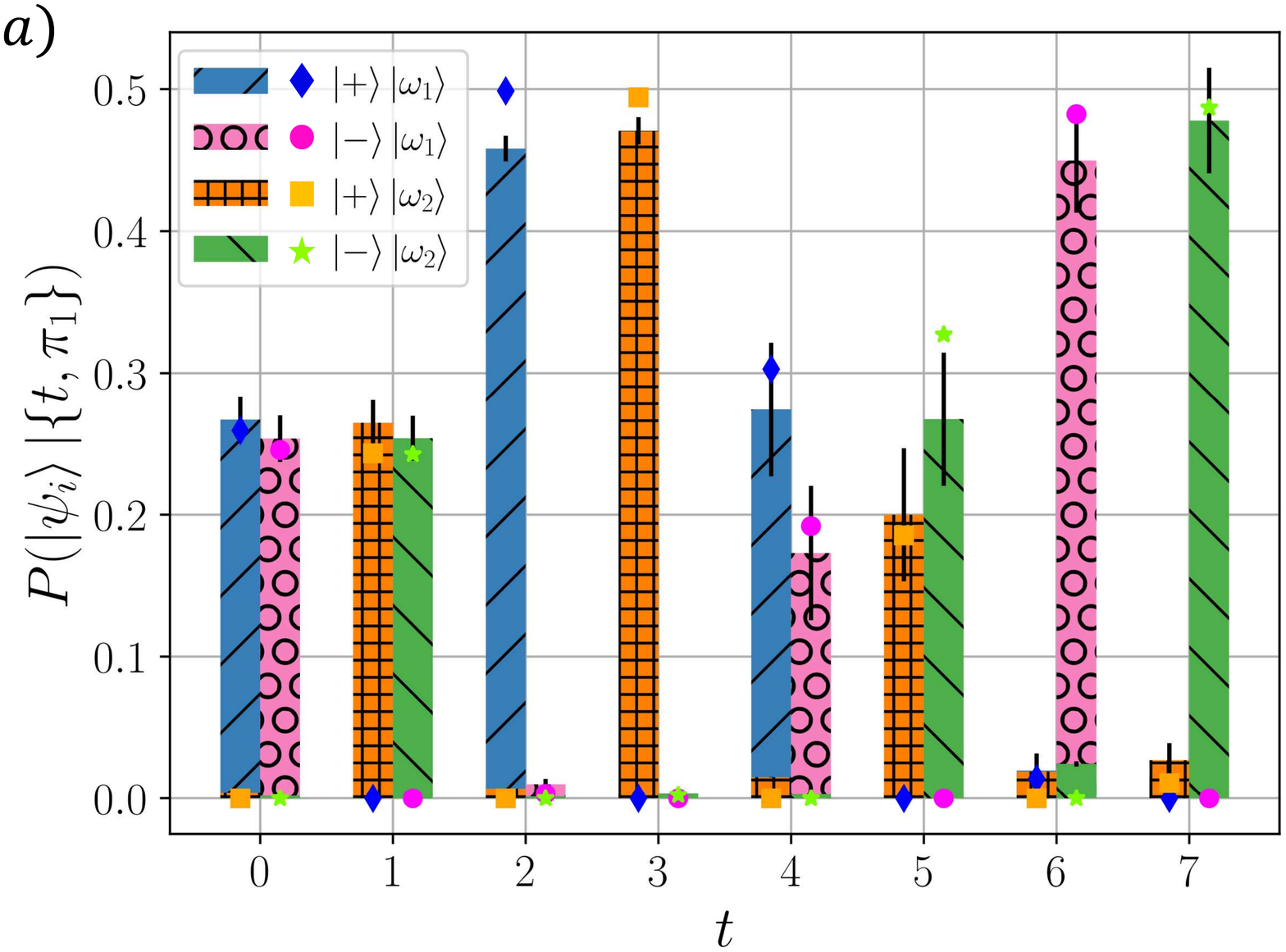}
       \label{}
       \end{subfigure}
        \begin{subfigure}{0.5\textwidth}
        \includegraphics[width=1.\columnwidth]{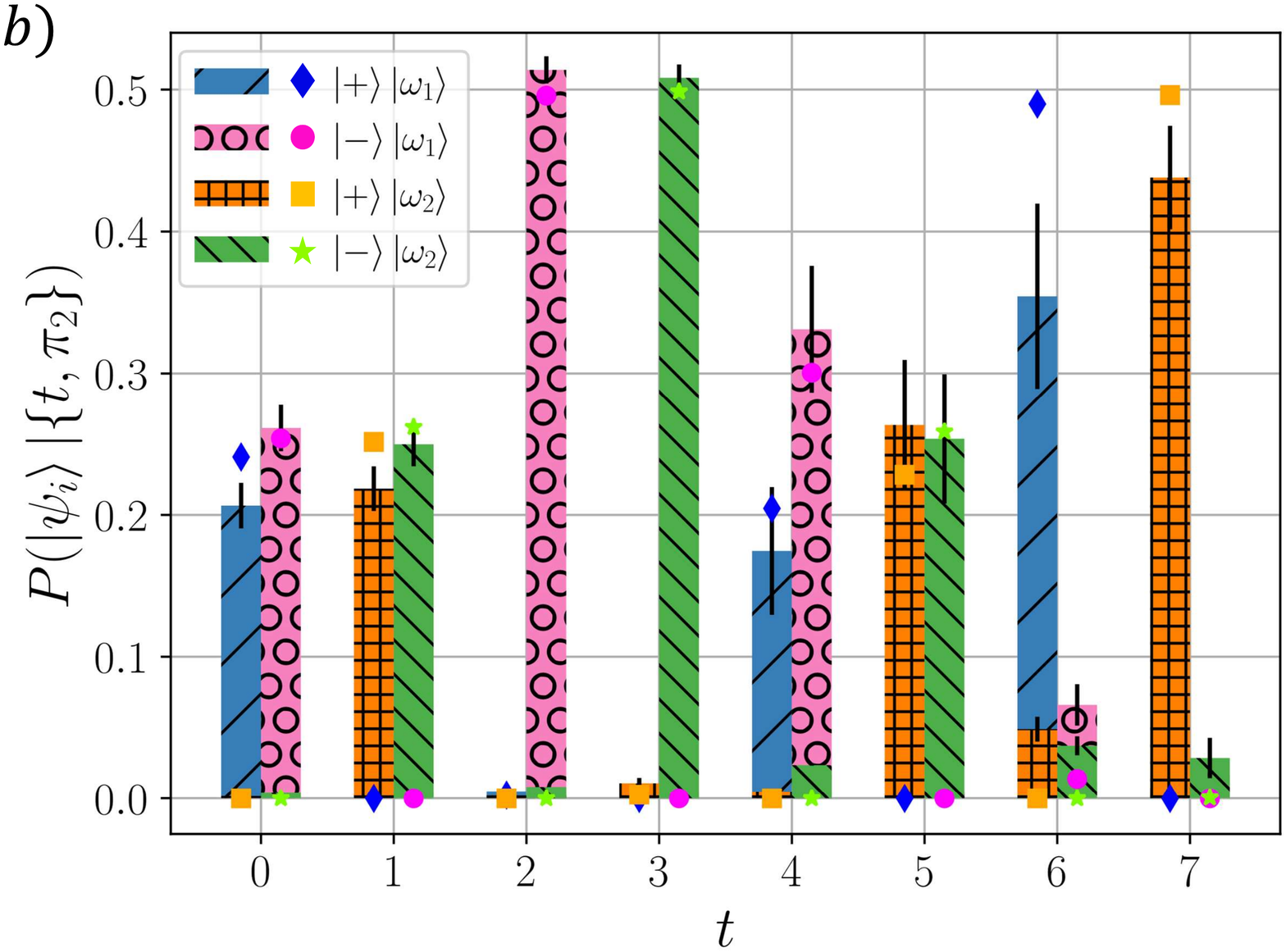}
        \label{}
       \end{subfigure}
        \begin{subfigure}{0.5\textwidth}
        \includegraphics[width=1.\columnwidth]{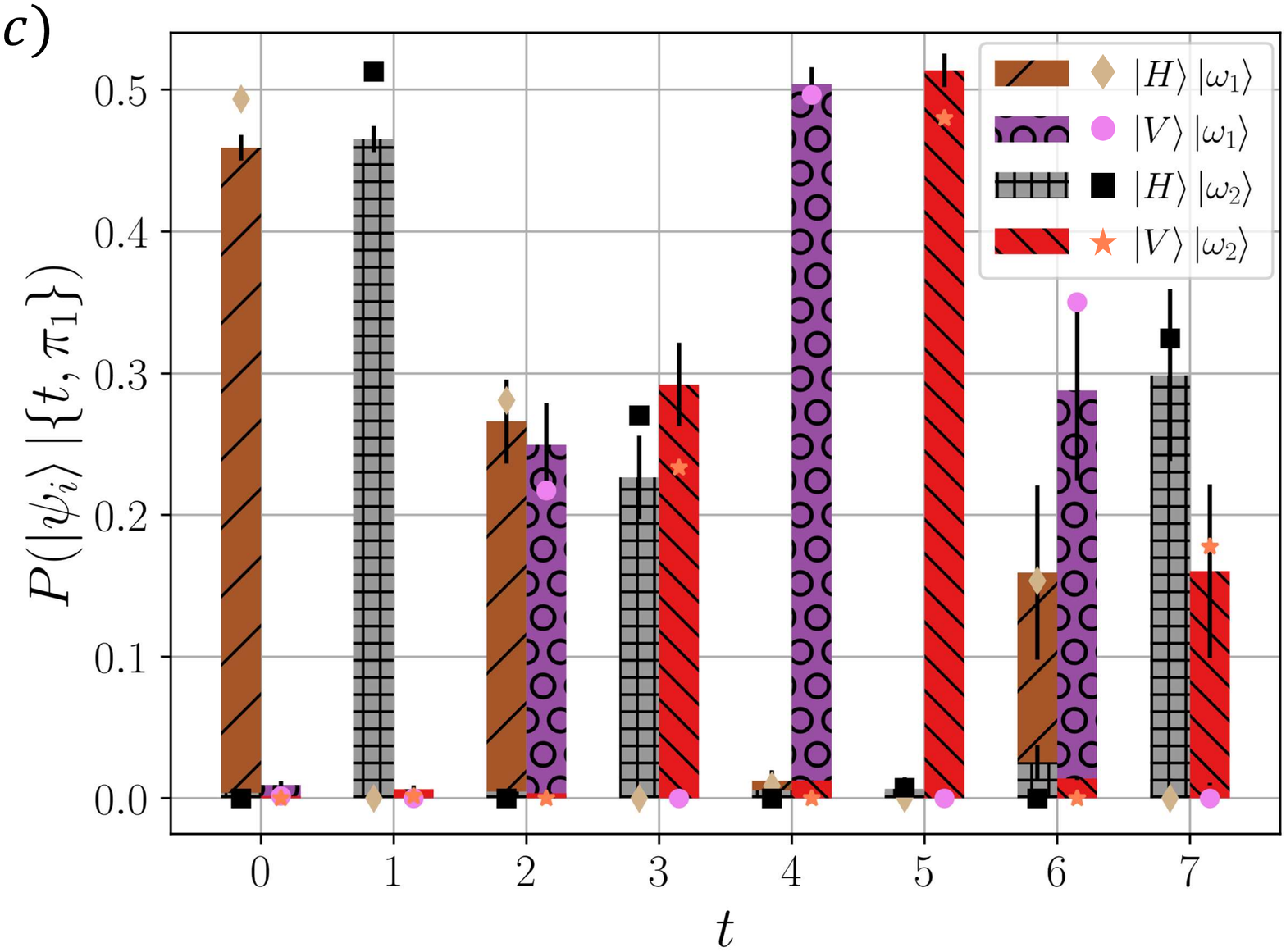}
        \label{}
       \end{subfigure}
        \begin{subfigure}{0.5\textwidth}
        \includegraphics[width=1.\columnwidth]{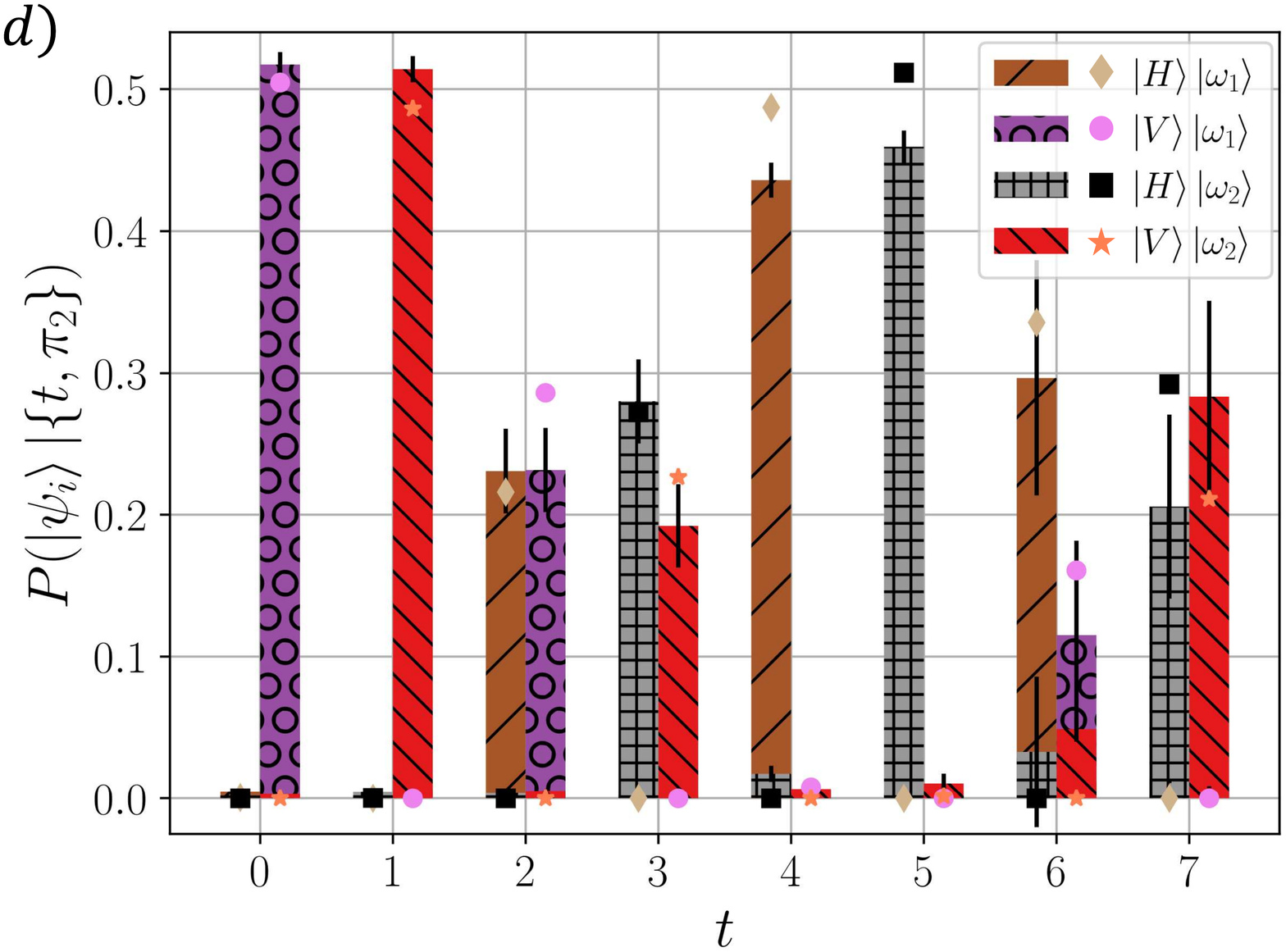}
        \label{}
       \end{subfigure}
        \caption{\textbf{Probabilities for each of the eight states conditioned to the detection of a photon in each time bin and in the a)-c) $\ket{H}$ polarization or b)-d) $\ket{V}$ polarization.} 
        \textit{ The data are normalized so as to have $P(\{t,\pi\})=1$ for any $\{t,\pi\}$. They are compared to their expected values, that are the results of a simulation run accounting for experimental parameters, here represented by symbols of similar colours with respect to their experimental counterpart. Some discrepancies are present, that are extensively discussed in the Methods section, together with the error computation. For each state, it is possible to recognize a middle-maximum-middle-minimum dynamics with a different time shifting. Some asymmetries can be ascribed to the slight unbalanced behavior of the Beam Splitter with respect to polarization. 
        } }
        \label{fig:results}
\end{figure*}

The time-multiplexing scheme we described above can be implemented by a linear optics setup, employing simple components and two photodetectors only, connected to a time tagger.\\
We encode the input states in the polarization and frequency of single photons generated by a QD source, driven under resonant two-photon excitation by a Ti:Sapphire laser at frequency $781.2\,nm$  to produce two photons at different wavelengths via the biexciton-exciton cascade. This approach produces photons with extremely high single photon purity \cite{hanschke2018quantum}. The setup can be divided in two parts: the first one, in Fig. \ref{fig:setup}a), is the single photon source, while the second one (Fig. \ref{fig:setup}b)) is the actual receiver. Further details about the setup are provided in the Methods section.\\
The states in Eq. \ref{eq:states} are prepared as follows: the frequency is selected by sending to the receiver in Fig. \ref{fig:setup}b) the single photons generated by an exciton-to ground transition (exciton or X for brevity) or a biexciton-to-exciton one (biexciton or XX). As depicted in Fig. \ref{fig:setup}a) the pump laser backreflection is filtered out by an optical Volume Bragg Filter (VBG) filter. The VBGs feature a wavelength tunable narrow-band reflection and the tunability is obtained by changing the angle of incidence. Two other VBG filters are used to extract photons emitted from the QD at the two different frequencies. In particular, the exciton wavelength is $780.3\,\rm{nm}$ and the biexciton one is $ 782.3\,\rm{nm}$. The FWHM of the emission, obtained with a Gaussian fit of the peaks, are $22.3\pm6\,\rm{pm}$ for the exciton and $37.3\pm7\,\rm{pm}$ for the biexciton, while an individual VBG filter has a bandwidth $<60\,\rm{pm}$ and an extinction ratio $OD3$ as stated by the manufacturer. The filter bandwidth is much wider than the QD linewidth, which allows to collect a specific emission line avoiding filtering losses, while still effectively separating different emission lines.
Afterwards, photons prepared in the frequency subspace we desire are sent via single-mode fibers to the second part of the setup in Fig. \ref{fig:setup}b) where the polarization state is prepared by a linear polarizer and a waveplate (WP) set composed of the sequence of a Quarter-Wave Plate (QWP), a Half-Wave Plate (HWP), and a QWP. The WP sequence corresponds to prepare the polarization state by a unitary preparation operation $\hat{U}_P$ and the discrimination protocol begins.\\
The first evolution step, corresponding to the application of the unitary $\hat{U}_F$ bringing the system from the initial to the intermediate network layer, is practically performed by the same WP set as the preparation stage, as depicted in Fig. \ref{fig:setup}b). Photons find first on their path an unbalanced Beam Splitter (BS) with an average reflectivity $R\approx27.5\%$. This value is chosen to maximize the amount of signal collected through the process, while keeping the detected population in different time bins as homogeneous as possible (more details are given in the Methods section). If photons are transmitted, they travel a loop of mirrors in which their polarization state is evolved by other waveplates, so that they eventually impinge again on the BS with a different polarization state.
If the photon is reflected, another loop evolution begins and the polarization state changes in turn. In this way we are able to apply a dynamics on the single photon polarization degree of freedom. The unitaries $\hat{U}_F$ and $\hat{U}_B$ regulating the evolution are chosen as follows: the complete loop evolution, corresponding to $\hat{U}_F\cdot \hat{U}_B$, rotates the polarization state of $90^{\circ}$ on the Bloch sphere, along the $\{\ket{H},\ket{+}\}$ plane, so that each possible initial polarization state will have a maximum probability of being sent out of the loop in $\ket{H}$ or $\ket{V}$ state at a different time. Thanks to that, we can set the evolution parameters to maximize some selected $P(\{t_i,\pi_i\}|\ket{\psi}_i)$ and to optimize the polarization-wise discrimination. The frequency-wise discrimination happens after the extraction from the first optical loop, addressed as $L_1$: single photons encounter another VBG filter along their path, which is set to reflect the exciton transition ones at $\omega_2$.
Reflected photons travel a supplemental optical loop, called $L_2$, after which they impinge again on the VBG and are redirected on the same path as the $\omega_1$ photons, but with a delay in time, tailored to be half the travel time of the $L_1$.
In this way, we end up with doubled time bins, one for each possible frequency.
The signal distribution in time is represented in Fig. \ref{fig:results}b). This time-binning procedure is possible because of the short lifetimes of the QD emission: in the particular case of the QD used in this experiment we measured a lifetime of 22$\pm$4 ps for the XX transition and 41$\pm$2 ps for the X transition. This guarantees an emission time distribution after each laser pulse which is much shorter than the period of the pulse train, minimizing in this way the cross-talking between two subsequent pulses.\\
We measure the signal output of the first eight time bins, corresponding to four $L_1$ trips plus the $L_2$ doubling, both for the $\ket{H}$ and $\ket{V}$ polarization.
In Fig. \ref{fig:results}, we report histograms of the sampled probabilities $P(\ket{\psi_i}|\{t,\pi\})$, of having a certain input state $\ket{\psi_i}$ if the photon is detected in $\{t,\pi\}$. These quantities, from which we can estimate the performance of our receiver, are proportional to the probabilities $P(\{t,\pi\}|\ket{\psi_i})$ we considered in our optimization procedure, according to Bayes' rule:
\begin{equation}
P(\ket{\psi_i}|\{t,\pi\})=\frac{P(\{t,\pi\}|\ket{\psi}_i)*P(\ket{\psi}_i)}{P(\{t,\pi\})}    
\label{eq:bayes}
\end{equation}
where $P(\{t,\pi\})=\sum_{j=1}^8 P(\{t,\pi\}|\ket{\psi_j})*P(\ket{\psi_j})$.\\

The periodical dynamics is clearly visible, differently time shifted for each different input state, where the maximum probabilities are achieved at a different $\{t_i,\pi_i\}$ for each $\ket{\psi_i}$. As an instance, we consider the probability values for the state $\ket{\psi_1}=\ket{+}\ket{\omega_1}$, represented in Fig. \ref{fig:results}a) and b) by blue bars. In this case, the system features XX frequency, then we only consider even time bins (corresponding to transmission through the VBG of $L_2$). The detection of an event at $t=2$ and $\pi=\pi_1$ points out that the input state was $\ket{+}\ket{\omega_1}$ with a probability that is maximum time-wise and state-wise. In fact, the bars in Fig. \ref{fig:results} are normalized to have $P(\{t,\pi\})=1$ for any $\{t,\pi\}$, namely to frame the situation in which something has been received and information has to be deduced from that. So, if we consider the input states to be equally likely, we see by Eq. \ref{eq:bayes} that the probability distributions in figure are just proportional to the $P(\{t,\pi\}|\ket{\psi}_i)$ that reflect the evolution of the system depending on its initial state. Indeed, the time-wise pattern middle-maximum-middle-minimum of probability values, that is clearly noticeable in Fig. \ref{fig:results}a), represents the fixed polarization rotation imposed by passage through $L_1$. 
While the time evolution is the same, the starting point changes, resulting in a probability dynamics that is shifted in time according to the initial state. A comparison between Fig. \ref{fig:results}a) and c) (i.e. population at the $H$ polarized output) yields exactly this behavior: initial polarization state $\ket{+}$ produces a dynamics that is shifted of two time bins  (corresponding to one passage through $L_1$) with respect to an initial $\ket{H}$ or $\ket{V}$. This is true for both the possible frequencies, so that, for example, the state $\ket{\psi_5}=\ket{+}\ket{\omega_2}$ has a time-wise probability which is identical to the one corresponding to $\ket{\psi_1}$, but for a single time bin shift (orange bars in Fig. \ref{fig:results}a)).
At the same time, the dynamics in the $V$ polarized output (reported in Fig. \ref{fig:results} b) and d)) is complementary to the $H$ polarized one.
Therefore, by measuring eight time bins, we are actually considering two maxima for each input state, one on the horizontal polarization and one on the vertical one.
The experimental data are shown in comparison with the expected ones, represented as scattered symbols, produced by a simulation performed taking into account experimental parameters. Indeed, the optical components feature some asymmetries that affect the protocol performance, the most relevant of which is the uneven response to polarization of the unbalanced BS. The reflectance and transmittance of the BS depend on the polarization of incoming light in such a way that, according to our analysis, the experimentally achievable average probability of correct guess is slightly lower than the theoretical optimum.
Moreover, these asymmetries can apparently yield an over-performance of the protocol for some  input states: for instance, in Fig. \ref{fig:results} d), the expected maximum guessing probability for the state $\ket{V}\ket{\omega_1}$ is higher than the theoretical maximum $0.5$. Indeed, this effect is compensated by under-performance for other states, so that the two contributions compensate when considering the average probability of correct guess, hence the average performances of the receiver.  

A thorough discussion of the setup imperfections is reported in the Methods section.\\
From the experimental samples, we can compute the average probability of correct guess, where we guess the input state to be $\ket{\psi}_i$ if the system is detected in the corresponding $\{t_i, \pi_i\}$, weighing with the total probability for the system to be found in $\{t_i,\pi_i\}$:
\begin{equation}
\bar{P}_{guess}=\frac{1}{2}\frac{1}{8}\sum_i P(\ket{\psi_i}|\{t_i,\pi_i\})P(\{t_i,\pi_i\})
\end{equation}
where we consider equally likely input states (hence the $1/8$ factor), and we sum two terms for each state, as mentioned above (hence the $1/2$ factor).
The result of this computation delivers a $\bar{P}_{guess}=0.486\pm0.002$ which is in agreement with the expected value $\bar{P}_{guess}^{sim}=0.488$, and it is also very close to the ideal performance $P_{guess}=0.5$.
The experimental errors were computed through a Montecarlo procedure based on the uncertainty over the experimental parameters. More details are reported in the Methods section.

\section*{Discussion}

We employed a solid state source, featuring excellent efficiency and control over different aspects of the emission process, to experimentally realize a scheme for discrimination among eight quantum states of actual single photons, encoded in their polarization and frequency degrees of freedom.
The protocol was implemented through a time-multiplexing dynamics that allows the achievement of optimal results using linear optics, with no feedback mechanisms and only two photodetectors.\\
The importance of our results resides not only on the fact that, to the best of our knowledge, this is the largest alphabet ever implemented with a single photon platform for QSD, but also from the novel employment of photon energy as an encoding variable. 
Energy is very robust to environmental noise and, in principle, it can even be used to span spaces with dimension $D>2$.\\
The possibility of writing information into photon frequency with such reliability derives from the exceptional properties of our solid state source: not only the source is nearly deterministic, but the generated photons also feature a convenient trade-off between fast emission and narrow bandwidth, granting reliable encoding and decoding effectiveness. 
There are, naturally, some limits to additional developments with this approach: the preparation of a superposition state in frequency as well as its possible manipulation inside the receiver are not trivial.
However, some recent efforts have been successfully devolved to efficient frequency conversion and manipulation \cite{manurkar2016multidimensional,kasture2016frequency,otterstrom2021nonreciprocal}. \\
Moreover, we deliver a novel experimental method for frequency-wise delay lines, that may feature a wide applicability.
For these reasons, we believe that our work can represent a relevant step in the implementation of high-dimensional quantum communication protocols and the start for further promising investigations. 

\section*{Methods}

\subsection*{Setup details}
The source of single photons is a single GaAs/AlGaAs placed in a circular Bragg resonator (CBR) cavity (also know as bullseye cavity) \cite{davanco2011circular}. The sample is grown with the Al-droplet etching technique on a GaAs commercial wafer \cite{huo2013ultra}. Then, the sample undergoes several processing steps to fabricate positioned and size-tailored CBR around selected QDs, following similar procedures to those previously developed for the GaAs/AlGaAs system \cite{liu2019solid, liu2017cryogenic}.
The cavity is designed to have the main resonant mode in the vicinity of the X-XX energies. In this way the lifetimes of the excitations are strongly reduced, with Purcell factors up to 6.
The sample is then placed in a cryostat at low temperature (5K) and a single QD is excited by focusing laser light with an aspheric lens placed inside the cryostat in a confocal configuration, see Fig. \ref{fig:setup}a). The signal from the QD is collected by the same lens and directed to the optical table. 
The XX level of the QD is resonantly excited by tuning the energy of a Ti:Sapphire pulsed laser at half of the transition energy from the ground state inducing a two-photon absorption. To enable the absorption, the QD charge environment must be stabilized with the aid of an uncollimated halogen lamp with blackbody spectrum \cite{nguyen2013photoneutralization,huber2016effects}.
After excitation, the QD relaxes to the ground state by emitting a cascade of two photons, correlated in polarization, at two different energies, with an energy difference that depends on the binding energy of the XX state (here equal to 2 nm). 
The pump pulses have a $80\,\rm{MHz}$ repetition rate, meaning that the narrowest time interval between two consecutive photon generations is around $12.5\,\rm{ns}$. We tailored the length of the optical loops in order to fit eight detection time bins in the $12.5\,\rm{ns}$ laser repetition period. To this aim, the two loops were set to $l_{L_1}=90\,\rm{cm}$ and $l_{L_2}=45\,\rm{cm}$.
As a consequence, a very narrow time distribution of the photon generation is required from the Quantum Dot, that is due to the fast recombination of GaAs/AlGaAs nanostructures further enhanced by Purcell effect, leading to the lifetime values  mentioned in the main text..
 The $L_2$ optical loop, also because of its relatively short length, must be aligned very carefully: the VBG reflects only a small band around a certain wavelength depending on the angle of incidence with a very high angular selectivity (circa $0.4\,\rm{mrad}$). Thus, in order to close the loop, it is necessary not only to impinge on the VBG with the same angle at the second reflection, but also to hit the same spot in order to align with the trasmitted beam. \\
To make the whole system flexible to wavelength changes, for example when using a different QD, the VBG is placed on a rotating mount that allows for a precise and flexible angular alignment. To adjust to different angles of reflection of the VBG filter the optical loop $L_2$ is placed on a platform that can rotate independently from the VBG filter.

\subsection*{Data analysis}
Measurements were carried over the eight time bins at the same time, with a count rate oscillating around $\approx3-3.5\cdot10^3/\rm{s}$, each measurement lasting $3$ minutes.
Before processing, data were cleaned off background noise, due to dark counts of the SPADs, continuous wave signal from the QD due to white light illumination, and environmental noise.\\
In Fig. \ref{fig:setup}a) we show the single photon source and part of the preparation stage, that consists in a wavelenght-wise post-selection. The biexciton photons have wavelenght $\lambda_1=780.27\,\rm{nm}$, while the exciton ones have $\lambda_2=782.25\,\rm{nm}$. 
A limitation of this approach arise when the linewidth of the photons is larger than the spectral bandwidth of the VBG filter. For the QD we exploited in this work a low intensity band in the vicinity of the X emission was transmitted through the VBG filter (most probably due to a non perfect alignment of the central energies of the VBG selecting the X line on the optical table and the one on the $L_2$ loop). To reduce this effect we placed two additional VBG filters on the optical table to rule out any contribution around the X line. 
The residual parasite counts are visible in Fig. \ref{fig:results}, especially for larger delays where the lower signal is more affected by noise and can be ascribed to the finite extinction ratio of the VBG ($OD3$). From the counts, we experimentally estimate a total extinction ratio of $0.0125$ for H polarization and $0.0055$ for V polarization corresponding to the double reflection on the VBG in $L_2$. This discrepancy can be straightforwardly attributed to an imperfect angular positioning of the VBG filter and, especially, an imperfect angle of impingement of photons after traveling the $L_2$ loop. Indeed, the setting of the latter is very difficult to achieve with a high precision and we estimate that an error of $0.1\%$ on the angle of incidence on the VBG can result in a decrease of one order of magnitude in the extinction ratio.
We don't include this effect in our statistical error computation, since, as we show below, it does not significantly affect the performance of the receiver.\\
Indeed, the most relevant imperfection, in terms of effects on the performance, is the previously mentioned asymmetrical reflectivity with respect to polarization of the BS in $L_1$: specifically, it features $R_H=0.26\pm0.02$ and $R_V=0.29\pm0.02$ together with losses of $\approx21\%$, that sum up with another $11\%$ of losses for every loop completion. 
Since the BS and the loop $L_1$ are travelled repeatedly, the asymmetry and the losses are very influential on the performance of the receiver, resulting on the suboptimal expected $P_{guess}$ value of $0.488$ instead of the ideal $0.5$. This estimate is in excellent agreement with our analysis of the experimental data, therefore confirming that the BS imperfection is the main limiting factor.\\
The experimental errors employed to compare the experimental data with the expected ones have been computed by considering possible errors in the parameter evolution and evaluation: indeed, the waveplates angles positioning and calibration are subject to random errors, that we estimated jointly as $\pm 1^{\circ}$ uncertainty on their angular position, and the BS reflectivities feature the uncertainty reported above. The errors are computed through a Montecarlo procedure and then summed in quadrature to the Poissonian error computed on the raw counts.\\
Without considering the losses, we are able to collect approximately the $98.5\%$ of the input signal by measuring the output of the first four passages through the $L_1$ loop (corresponding to the first eight time bins). Therefore, the scheme can perform both very efficiently and effectively, once losses are reduced.

\section*{Acknowledgments} We acknowledge support from MIUR (Ministero dell’Istruzione, dell’Università e della Ricerca) via project PRIN 2017 “Taming complexity via QUantum Strategies a Hybrid Integrated Photonic approach” (QUSHIP) and via the funding program SEED PNR 2021. This project has received funding from the European Union’s Horizon 2020 Research and Innovation Program under Grant Agreement no. 899814 (Qurope) and from the QuantERA II Programme that has received funding from the European Union’s Horizon 2020 research and innovation programme under Grant Agreement No 101017733 via the project QD-E-QKD.

%
%
%
%
%
%
%
%
\end{multicols}

\bibliography{biblio.bib}

\end{document}